\documentclass[1p]{elsarticle}

\usepackage{lineno, hyperref, amsfonts, amsmath, graphicx, wrapfig, natbib, fancyhdr}
\graphicspath{{./images/}}

\journal{under review}

\setcitestyle{elsarticle-num}

\begin{document}
\begin{frontmatter}
	\title{Toward the biological model of the hippocampus as the successor representation agent}
	\author{Hyunsu Lee\corref{cor1}}
	\address{Department of Anatomy, School of Medicine, Keimyung university, Daegu, Republic of Korea}
	\ead{neuroana@dsmc.or.kr}
	
	\cortext[cor1]{Corresponding author}
	
	\begin{abstract}
		
		The hippocampus is an essential brain region for spatial memory and learning. Recently, a theoretical model of the hippocampus based on temporal difference (TD) learning has been published. Inspired by the successor representation (SR) learning algorithms, which decompose value function of TD learning into reward and state transition, they argued that the rate of firing of CA1 place cells in the hippocampus represents the probability of state transition. This theory, called predictive map theory, claims that the hippocampus representing space learns the probability of transition from the current state to the future state. The neural correlates of expecting the future state are the firing rates of the CA1 place cells. This explanation is plausible for the results recorded in behavioral experiments, but it is lacking the neurobiological implications.

		Modifying the SR learning algorithm added biological implications to the predictive map theory. Similar with the simultaneous needs of information of the current and future state in the SR learning algorithm, the CA1 place cells receive two inputs from CA3 and entorhinal cortex. Mathematical transformation showed that the SR learning algorithm is equivalent to the heterosynaptic plasticity rule. The heterosynaptic plasticity phenomena in CA1 were discussed and compared with the modified SR update rule. This study attempted to interpret the TD algorithm as the neurobiological mechanism occurring in place learning, and to integrate the neuroscience and artificial intelligence approaches in the field.
	\end{abstract}
	

\end{frontmatter}


\section{Introduction}
	The navigating environment, including foraging, nesting, and finding a mate, is the essential ability for surviving animals. To achieve a navigating goal, animals developed a specific nervous system, which is believed to be the hippocampus. After discovery of place cell \citep{okeefe1971}, it emerges the cognitive map theory which believed that hippocampus encodes spatial memory \citep{mcclelland1995}. With the case studies of patient H.M. \citep{milner1968}, the cognitive map theory has expanded into a complementary learning system \citep{kumaran2016}. The hippocampus encodes and transfers not only spatial memory but also semantic memory and episodic memory to the neocortex. Episodic memories linking specific locations and events are essential to animal survival. It is important to remember not only the location of the foods and the predators, but also its existence. From the point of view of reinforcement learning (RL) in psychology, we can regard foods and predators as rewards and punishments, respectively. However, it is not well known how the hippocampus processes the informations of rewards and punishments for spatial and episodic memories.
	
	Leveraging the concept of RL derived from psychology is not limited to neuroscience. In artificial intelligence, they have conducted research using the concept of RL for a long time \citep{sutton2018}. The recent research results, such as AlphaZero and AlphaStar, have shown to exceed human performance in playing Go and StarCraft II \citep{silver2017, vinyals2019}. From a computational point of view, RL computes the value of the current state or environment and predicts the value of the future state after the agent takes action. To achieve this computational goal, artificial intelligence researchers have developed many algorithms. Among them, temporal difference (TD) learning is related to neuroscientific mechanisms of RL found in the brain \citep{schultz1998, suri2002}. TD learning updates the expectations for the value of the current state by observing the difference between the predicted and observed values after taking the action: this is the TD error. It is equivalent to prediction error in the concept of the predictive coding \citep{odoherty2003}. Although the prediction error hypothesis gives integrating insight for psychology, artificial intelligence, and neuroscience, it is little known about the biological mechanisms of how learning of spatial memory in the hippocampus is occurring by prediction errors.
	
	One limitation of TD learning is that the agent does not differentiate between the rewards earned and the states change when it updates the value of the states after taking an action. Because of that, even if the position of the reward changes in the same environment, the TD agent has to learn again from the scratch. But animal learning is more robust and adaptive than this. To overcome this problem, successor representation (SR) learning has been proposed \citep{dayan1993}. This algorithm decomposes the reward and state transition functions from the value function update. By applying this idea to hippocampal learning, a recently published study developed predictive map theory \citep{stachenfeld2017}. The study reported that the learned results from exploring the environment according to the SR learning rule were similar to the responses recorded in the animal hippocampus. They argued that place cell of the hippocampus represents the probability of transition the successor location from the current location rather than a neural correlate of simple geodetic. They, however, have not explored the biological mechanisms of SR learning.
	
	In this article, we will modify the SR model by the biological basis of the hippocampus. To compensate for the lack of biological mechanisms in the SR model, we will derive the synaptic weight update rule from the SR learning rule. We will treat the term representing states in the SR model as presynaptic neuron of the CA1 place cells. As a result, we reveal that the co-activity of the two presynaptic inputs to the CA1 place cell is important to update the synaptic weights. And it is noteworthy that this update rule is comparable to the heterosynaptic plasticity of the CA1 place cells.

\section{Key ideas from the SR learning model}

	The key idea of predictive map theory is that place cells encode expectations of future location from a viewpoint of the current location: the firing rate of the place cells reflects the probability of transition \citep{stachenfeld2017}. If the place cell encodes a location close to the animal's current location, it means that the transitional probability is high, therefore the firing rate of the place cell will also be high. It is noteworthy that the transitional probability depends not only on the simple Euclidean distances but also on the choices of the agent. Even if the location is far away from the current location, if there is a reward, the transitional probability will be high, and vice versa. Besides reward/punishment, the habitual behavior of the agent, obstacles, and detours also affects the probability of transition. The predictive map theory considers these factors. But the classical cognitive map theory does not consider above factors but considers only the Euclidean distance.
	
	The \cite{stachenfeld2017} introduced the SR learning model as an algorithm to elaborate predictive map theory. The SR learning was derived from the TD algorithm that discounts the value of a state depending on the distance to the reward. (From now on, I will use the term state instead of location. The term state is a term mainly used for reinforcement learning and includes not only a spatial location but also an agent state, including selectable actions.) The expectation of distal rewards $R(s_{t})$ multiplied by a discount factor $\gamma \in [0,1]$ given current state shows the value of state $V(s)$ 
	\begin{equation} \label{value_vanila}
	    V(s) = \mathbb{E}[\sum^{\infty}_{t=0} \gamma^{t}R(s_{t}) | s_{0} = s ]
	\end{equation}
	$s(t)$ indicates the state visited at time $t$.
	
	The essential idea of the SR learning is that the value function is a combination of prediction of transition to the successor state $s'$ and the reward of that state \citep{dayan1993}. Thus, we can decompose them as the following equation.
	
	\begin{equation} \label{value_decomposed}
		V(s) = \sum_{s'}M(s, s')R(s')
	\end{equation}
	
	We can derive the function $M(s, s')$ from the value function (\ref{value_vanila}) by imputing trajectories of the states instead of $R(s_{t})$ as the following equation.
	\begin{equation} \label{Meq}
		M(s, s^{'}) = \mathbb{E}[\sum^{\infty}_{t=0} \gamma^{t} \mathbb{I}(s_{t} = s')  | s_{0} = s ]
	\end{equation}
	$\mathbb{I}(s_{t}=s')$ is a Boolean function that returns 1 if the agent visited $s’$ at time $t$, or 0 otherwise.
	
	The agent can learn the environment incrementally by the TD algorithm. After taking action and observing the next state, it updates the value of the current state. Therefore, we can update the estimation of $M(s, s')$ by same manner as the following equation. 
	\begin{equation} \label{update_M}
		M_{t+1}(s_{t}, s') = M_{t}(s_{t}, s') + \eta[ \mathbb{I}(s_{t} = s') + \gamma M_{t}(s_{t+1}, s') - M_{t}(s_{t}, s') ]
	\end{equation}
	
	Although the updating TD algorithm manipulates only the information of the next state, it is the bootstrap method based on previously learned values. Thus, as defined by equation (\ref{Meq}), the horizon of $s'$ in $M(s, s')$ can be infinite. Mathematically, it is appropriate for the agent to initialize the matrix M to the identity matrix; the $M(s, s')$ is 1 if $s = s'$, otherwise 0. This is because if the agent does not take action without prior knowledge of the environment, the only successor state of the current state is the current state. The size of row and column of matrix $M$ must be equal to the number of the states in the environment that the agent will explore (I will discuss this in the limitations). 
	
	The simulation results of the SR agent shown in the abovementioned paper \citep{stachenfeld2017} are interesting. The estimated $M(s, s')$ value after learning seems similar to the firing pattern of place cells recorded in the hippocampus CA1 of the murine model. In the meantime, the simple geodesic model has not fully explained the recorded data in the murine model. It cannot reflect the dynamic response of place cells to environmental changes such as barriers or detour, or to the direction of the moving animal. These diverse and dynamic environmental changes that are likely to occur in the real world, however, change the transitional probability ($M(s,s')$ value) of the SR model. For examples, on the one-way track,  place fields often show the ramp patterns that gradually increase and abruptly decrease, which are asymmetric firing patterns. Considering the SR model, the transitional probability gradually increases before the animal arrives at the place field. Since the animal on the one-way track rarely goes back, the transitional probability to the place field after passing it is almost zero. Therefore, the SR model explains the firing pattern of the place cells better than the simple geodesic model.

\section{Limitations of the SR model}
	Based on the similarity of the pattern between the estimated $M$ and the place field, we can interpret the firing pattern of place cells as the transitional probability to the future state. Despite the similar generated pattern, however, the SR model has some limitations from the perspective of biological neurosciences.
	
	The major concern with the SR model is that it did not reflect realistic interactions between biological agents and the environment. In the SR model, the artificial intelligence (AI) agent must determine the size of the $M$ matrix based on the number of movable points in the environments to be exposed in prior: it is given by the human programmer. If we expose the AI agent to a new environment after learning the other environment once, the agent must not only re-learn the transitional probability in the new environment but also change the size of the $M$ matrix according to the new environment. Knowing the number of place fields in an unfamiliar environment in prior, however, is not a plausible condition for biological agents. Rather, they generate the place fields by exploring the unfamiliar environment; it is impossible for the AI agent of the SR model.
	
	Despite of the differences, the similarity of the response patterns between $M$ and place cells makes plausible reasoning for predictive coding in the hippocampus. In order to apply predictive map theory to interpret data from animal, we must interpret the algorithmic level description of the SR model at the implementation level of the biological brain, especially at the cellular level.
	
	The SR model agent observes the result of the taken action; and it compare $M_{t}(s_{t},s')$ with $M_{t}(s_{t+1},s')$ to update the $M$ matrix (equation (\ref{update_M})). This is a simple arithmetic at the algorithm level, but not at the cellular level. After visiting the next place field, the place cell should update the firing rate for the current place field by according to the observed result. Based on this assumption, we can raise some biological questions. How the firing rate of place cell for the same sensory input (same place field) changes: what is the fundamental mechanism for that? If changing the synaptic weights between neurons is the answer, what is the update rule for the synaptic plasticity? And how the neuron perceive the temporal difference of the inputs from the current and the next place filed? To apply biological analysis, we should address these issues in the SR model.	

\section{Toward biological model}
	
	\begin{figure}[h]
		\centering
		\includegraphics{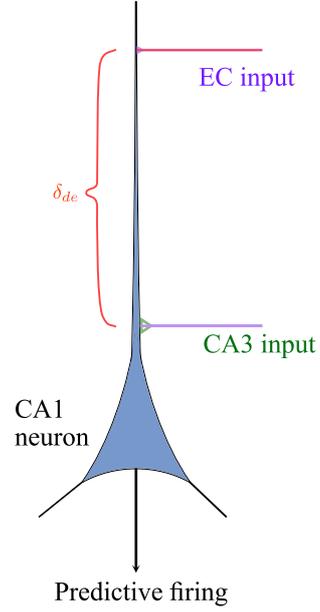}
		
		\caption{Schematic figure for CA1 place cells and its presynaptic inputs.}
		\label{fig:CA1}
	\end{figure}
	
	Given that the $M_{t}(s_{t},s')$ reflects the activity of place cells, we can suppose it as a composition of synaptic weight and presynaptic neural activity as the following equation.
	\begin{equation} \label{synaptic_M}
	    M_{t}(s_{t}, s') = W_{t} \cdot Pre(s_{t})
	\end{equation}
	where $W_{t}$ and $Pre(s_{t})$ is a vector of synaptic weights and of presynaptic activities at the time step, respectively. The presynaptic activities mean sensory inputs from other neural layers or brain regions; $Pre(s_{t})$ acting like $\mathbb{I}(s_{t} = s')$ in equation (\ref{Meq}) estimates the occupancies of the individual states \citep{vertes2019}. Thus, we can presume that the inner product of $W_{t}$ and $Pre(s_{t})$ shows the estimated values related to the transitional probability, which is equivalent to the place cell activities. For the convenience of the discussion, we omitted the activation function of the artificial neural function approximation. 
	
	To derive the synaptic update rule for biological neuron from equation (\ref{update_M}) of the SR model, we can replace the $M(s,s')$ with $W \cdot Pre(s)$ as following.
	
	\begin{equation} \label{hetero_syn_vanila}
		W_{t+1} \cdot Pre(s_{t}) = W_{t} \cdot Pre(s_{t})  + \eta[ Pre(s_{t}) + \gamma W_{t} \cdot Pre(s_{t+1}) - W_{t} \cdot Pre(s_{t}) ]
	\end{equation}
	
	Note that we still should deal with the presynaptic activities occurring at $S_{t}$ and $S_{t+1}$ simultaneously: for animals, $S_{t}$ is the place field just before, and $S_{t+1}$ is the place filed just observed. To resolve this issue biologically, we can hypothesize that two distinct presynaptic neural layer convey information of states, $S_{t}$ and $S_{t+1}$, simultaneously. Two major presynaptic inputs to the CA1 place cells are Schaffer collateral inputs from the CA3 and temporal-ammonic inputs from the entorhinal cortex (EC)(Figure \ref{fig:CA1}). 
	Thus, we can rewrite equation (\ref{hetero_syn_vanila}) as follows.
	
	\begin{equation} \label{hetero_M}
	\begin{split}
		W_{t+1} \cdot Pre^{CA3}(s_{t}) = & W_{t} \cdot Pre^{CA3}(s_{t})  +	\\	
		 & \eta[ Pre^{CA3}(s_{t}) +  \gamma W^{EC} \cdot Pre^{EC}(s_{t+1}) - W_{t} \cdot Pre^{CA3}(s_{t}) ]
	\end{split}
	\end{equation}
	where $W_{t}$ and $W^{EC}$ are synaptic weights for CA3 inputs and for EC inputs, respectively. $Pre^{CA3}$ and $Pre^{EC}$ are presynaptic activities of the CA3 and of the EC, respectively. Here, it is assumed that CA3 and EC encode information of $S_{t}$ and $S_{t+1}$, respectively, considering the delay caused by the hippocampal trisynaptic circuit \citep{dudman2007, lisman1999}. 
	
	The change of CA1 firing response for the place field $S_{t}$ occurs by updating the synaptic weights from CA3 inputs. Based on equation (\ref{hetero_M}), we can derive the synaptic weight update rule as following.
	\begin{equation} \label{hetero_update}
	\begin{split}
		\Delta W \cdot Pre^{CA3}(s_{t}) & =  \eta[ Pre^{CA3}(s_{t}) + \gamma W^{EC} \cdot Pre^{EC}(s_{t+1}) - W_{t} \cdot Pre^{CA3}(s_{t}) ] \\
        \Delta W &= \eta[\gamma W^{EC} \cdot Pre^{EC}(s_{t+1}) \cdot inv(Pre^{CA3}(s_{t})) - W_{t} + 1] 
	\end{split}		
	\end{equation}
	where $inv(Pre^{CA3})$ is the Samelson inverse of $Pre^{CA3}$ vector. 
	
	In our derived update rule, $\Delta W$ depends on the dot product of $Pre^{EC}$ and $Pre^{CA3}$. Biologically, this implies that simultaneous activation of the CA3 and the EC enhances the synaptic weight from CA3 input to the CA1 place cells. Modulation of the synapse by cooperative activity of two distinct presynaptic inputs is called the heterosynaptic plasticity. Deriving equation (\ref{hetero_update}) from the SR model implies that the heterosynaptic plasticity occurring at the CA1 place cells is the candidate for the biological implementation of the predictive map theory, the SR learning model. 
	
\section{Discussion}
	In recent years, the convergences of the field of AI and neuroscience have brought new ideas to each other, which leads to interesting results. The TD learning idea gave new insights that interpret the CA1 recording results of a moving animal in spatial navigation task \citep{stachenfeld2017}. In the study, the authors argued that the recorded pattern of CA1 place cells is comparable to the transitional probability matrix of the SR model, and proposed the predictive map theory of the hippocampus. In this article, we integrated the SR model into the hippocampal anatomical architecture, and derived heterogeneous synaptic plasticity rules from the SR model.
	
	To support the predictive map theory, we assumed that the CA1 place cells require two distinct presynaptic inputs to update their firing rate with the heterosynaptic plasticity. The input-timing dependent plasticity (ITDP) is a good candidate for the biological implementation of synaptic weight modification for predictive coding \citep{dudman2007}. They found that preceded stimulus from the EC before stimulus from the CA3 potentiated the synaptic strength between CA3 and CA1 pyramidal neurons. When the stimuli came in the opposite order, they observed no change in both synaptic strength. Thus, they called this phenomenon as the ITDP. Based on the simulation, they presumed that the candidate mechanism for the ITDP is the potentiated $\text{Ca}^{2+}$ transients of proximal dendrite elicited by stimuli to distal dendrite from EC. The opposite order of stimuli did not sufficiently potentiate the $\text{Ca}^{2+}$ transients of the proximal dendrite. The propagating excitatory post-synaptic potential (EPSP) from distal dendrite was enough slow to be summated with the elicited EPSP at proximal dendrite where receive inputs from CA3. The summated EPSP opens the NMDA receptor, which is important for the influx of $\text{Ca}^{2+}$ starting the intracellular signal cascades for synaptic potentiation. 
	
	Conjunctive activation of EC and CA3 input leads not only to eliciting synaptic plasticity but also to forming a place field of moving rodent. Prior to forming new place fields, preceding dendritic plateau potential was observed in vivo experiment \citep{bittner2015}. This preceding dendritic plateau potential required an interaction between the EC and CA3 inputs at the proper time interval and sequence. Similar to the ITDP, the dendritic potential elicited $\text{Ca}^{2+}$ transients that induced potentiation of EPSP from CA3. Since the time window width of this asymmetric synaptic potentiation was a few seconds, they named the synaptic rule as behavioral time scale synaptic plasticity \citep{bittner2017}. This biological mechanism reported in CA1 pyramidal neurons is comparable to our heterosynaptic plasticity rule derived from the TD algorithm: to the best of my knowledge, this is a novel finding. 
	
	Based on the anatomy of the hippocampus, the major inputs to CA1 place cells can be divided into intrinsic input and extrinsic input. The extrinsic input from outside of the hippocampus conveys sensory information from multiple sensory cortical areas. Among them, grid cells in the EC project their axons to the distal dendrites of CA1 place cells via the temporal-ammonic pathway \citep{fyhn2004, hafting2005}. The grid cells are characterized by an increase in firing rate as an animal moves a certain distance in the environment, forming a hexagonal grid field. The hexagonal grid field acts similarly to the latitude and longitude of the map, providing an allocentric location for moving animals \citep{buzsaki2013}. Thus, the extrinsic input from the EC might provide the immediate location of the agent: it would provide information for $M_{t}(s_{t+1}, s')$ in the SR model. 

	Another component of the SR model, $M_{t}(s_{t}, s')$, can correspond to information from CA3 coming along the Schaffer collateral pathway, the intrinsic input to CA1 place cells. The CA3 receive signals simultaneously from the EC and the dentate gyrus where receive signals via perforant pathway from the EC. Together the mossy fiber input and EC input, the recurrent connection of CA3 is the key feature of regarding CA3 as an attractor network \citep{rebola2017}. The attractor network based on the SR model has been reported to resemble preplay and rapid path planning, which are key features of the hippocampus \citep{corneil2015}. Based on these anatomical features, there have been suggestions that CA1 would be a subregion for predictive coding that compares the outputs of CA3 with sensory information from the EC \citep{lisman1999}.
	
	In terms of reward learning, however, both original SR model and our model did not resolve where is the region of brain composing state transitions and reward functions together, and how the composition occurs. One of the projection targets of the hippocampus is the ventral striatum, where is believed to be forming value functions by integrating information from other brain regions  \citep{johnson2007, lansink2009, pennartz2011}. In relation to value functions, the orbitofrontal cortex is one region where reward-based decision-making occurs \citep{rushworth2011}. The hippocampal-prefrontal replay has been reported to be important for spatial learning and memory-based decision making \citep{shin2019}. Therefore, further research is needed on how each region, including hippocampus, ventral striatum, and orbitofrontal cortex, interacts and learns the value of a given environmental state \citep{hirokawa2019}.
	
	We have discussed integrating the biological findings and algorithmic level to solve the spatial navigation problem. The functions of the entorhinal cortex and hippocampus are not limited to representations of places, but are diverse in concept representation, semantic memory, and episodic memory \citep{buzsaki2013}. The discovery of these functions in the same anatomical structure suggests that they employ similar algorithms. We can expect further research to apply the biological implementation of the SR model to these cognitive functions \citep{momennejad2017}.
	
\section{Credit author statement}
	Hyunsu Lee: Conceptualiztion, Methodology, Writing, Funding acquisition
	
\section{Acknowledgements}
	This study was supported by the National Research Foundation of Korea(NRF) grant funded by the Korea government(MSIT; Ministry of Science and ICT)(No. NRF-2017R1C1B507279).


\bibliography{hippo_sr_net}

\begin{thebibliography}{29}
\providecommand{\natexlab}[1]{#1}
\providecommand{\url}[1]{\texttt{#1}}
\expandafter\ifx\csname urlstyle\endcsname\relax
  \providecommand{\doi}[1]{doi: #1}\else
  \providecommand{\doi}{doi: \begingroup \urlstyle{rm}\Url}\fi

\bibitem[Bittner et~al.(2015)Bittner, Grienberger, Vaidya, Milstein, Macklin,
  Suh, Tonegawa, and Magee]{bittner2015}
K.~C. Bittner, C.~Grienberger, S.~P. Vaidya, A.~D. Milstein, J.~J. Macklin,
  J.~Suh, S.~Tonegawa, and J.~C. Magee.
\newblock Conjunctive input processing drives feature selectivity in
  hippocampal ca1 neurons.
\newblock \emph{Nat. Neurosci.}, 18:\penalty0 1133, 2015.
\newblock \doi{10.1038/nn.4062}.

\bibitem[Bittner et~al.(2017)Bittner, Milstein, Grienberger, Romani, and
  Magee]{bittner2017}
K.~C. Bittner, A.~D. Milstein, C.~Grienberger, S.~Romani, and J.~C. Magee.
\newblock Behavioral time scale synaptic plasticity underlies ca1 place fields.
\newblock \emph{Science}, 357:\penalty0 1033--1036, 2017.
\newblock \doi{10.1126/science.aan3846}.

\bibitem[Buzsáki and Moser(2013)]{buzsaki2013}
G.~Buzsáki and E.~I. Moser.
\newblock Memory, navigation and theta rhythm in the hippocampal-entorhinal
  system.
\newblock \emph{Nat. Neurosci.}, 16:\penalty0 130--138, 2013.
\newblock \doi{10.1038/nn.3304}.

\bibitem[Corneil and Gerstner(2015)]{corneil2015}
D.~S. Corneil and W.~Gerstner.
\newblock Attractor network dynamics enable preplay and rapid path planning in
  maze--like environments.
\newblock In C.~Cortes, N.~D. Lawrence, D.~D. Lee, M.~Sugiyama, and R.~Garnett,
  editors, \emph{Advances in Neural Information Processing Systems 28}, pages
  1684--1692. Curran Associates, Inc., 2015.

\bibitem[Dayan(1993)]{dayan1993}
P.~Dayan.
\newblock Improving generalization for temporal difference learning: The
  successor representation.
\newblock \emph{Neural Computation}, 5:\penalty0 613--624, 1993.
\newblock \doi{10.1162/neco.1993.5.4.613}.

\bibitem[Dudman et~al.(2007)Dudman, Tsay, and Siegelbaum]{dudman2007}
J.~T. Dudman, D.~Tsay, and S.~A. Siegelbaum.
\newblock A role for synaptic inputs at distal dendrites: instructive signals
  for hippocampal long-term plasticity.
\newblock \emph{Neuron}, 56:\penalty0 866--879, 2007.
\newblock \doi{10.1016/j.neuron.2007.10.020}.

\bibitem[Fyhn et~al.(2004)Fyhn, Molden, Witter, Moser, and Moser]{fyhn2004}
M.~Fyhn, S.~Molden, M.~P. Witter, E.~I. Moser, and M.-B. Moser.
\newblock Spatial representation in the entorhinal cortex.
\newblock \emph{Science}, 305:\penalty0 1258--1264, 2004.
\newblock \doi{10.1126/science.1099901}.

\bibitem[Hafting et~al.(2005)Hafting, Fyhn, Molden, Moser, and
  Moser]{hafting2005}
T.~Hafting, M.~Fyhn, S.~Molden, M.-B. Moser, and E.~I. Moser.
\newblock Microstructure of a spatial map in the entorhinal cortex.
\newblock \emph{Nature}, 436:\penalty0 801--806, 2005.
\newblock \doi{10.1038/nature03721}.

\bibitem[Hirokawa et~al.(2019)Hirokawa, Vaughan, Masset, Ott, and
  Kepecs]{hirokawa2019}
J.~Hirokawa, A.~Vaughan, P.~Masset, T.~Ott, and A.~Kepecs.
\newblock Frontal cortex neuron types categorically encode single decision
  variables.
\newblock \emph{Nature}, 576:\penalty0 446--451, 2019.
\newblock \doi{10.1038/s41586-019-1816-9}.

\bibitem[Johnson et~al.(2007)Johnson, van~der Meer, and Redish]{johnson2007}
A.~Johnson, M.~A. van~der Meer, and A.~D. Redish.
\newblock Integrating hippocampus and striatum in decision-making.
\newblock \emph{Curr. Opin. Neurobiol.}, 17:\penalty0 692--697, 2007.
\newblock \doi{doi.org/10.1016/j.conb.2008.01.003}.

\bibitem[Kumaran et~al.(2016)Kumaran, Hassabis, and Mcclelland]{kumaran2016}
D.~Kumaran, D.~Hassabis, and J.~L. Mcclelland.
\newblock What learning systems do intelligent agents need? complementary
  learning systems theory updated.
\newblock \emph{Trends Cog. Sci.}, 20:\penalty0 512--534, 2016.
\newblock \doi{10.1016/j.tics.2016.05.004}.

\bibitem[Lansink et~al.(2009)Lansink, Goltstein, Lankelma, McNaughton, and
  Pennartz]{lansink2009}
C.~S. Lansink, P.~M. Goltstein, J.~V. Lankelma, B.~L. McNaughton, and C.~M.
  Pennartz.
\newblock Hippocampus leads ventral striatum in replay of place-reward
  information.
\newblock \emph{PLoS Biol.}, 7:\penalty0 e1000173, 2009.
\newblock \doi{doi.org/10.1371/journal.pbio.1000173}.

\bibitem[Lisman(1999)]{lisman1999}
J.~E. Lisman.
\newblock Relating hippocampal circuitry to function: recall of memory
  sequences by reciprocal dentate--ca3 interactions.
\newblock \emph{Neuron}, 22:\penalty0 233--242, 1999.
\newblock \doi{10.1016/S0896-6273(00)81085-5}.

\bibitem[McClelland et~al.(1995)McClelland, McNaughton, and
  O’Reilly]{mcclelland1995}
J.~L. McClelland, B.~L. McNaughton, and R.~C. O’Reilly.
\newblock Why there are complementary learning systems in the hippocampus and
  neocortex: insights from the successes and failures of connectionist models
  of learning and memory.
\newblock \emph{Psychological review}, 102:\penalty0 419--457, 1995.
\newblock \doi{10.1037/0033-295X.102.3.419}.

\bibitem[Milner et~al.(1968)Milner, Corkin, and Teuber]{milner1968}
B.~Milner, S.~Corkin, and H.-L. Teuber.
\newblock Further analysis of the hippocampal amnesic syndrome: 14-year
  follow-up study of hm.
\newblock \emph{Neuropsychologia}, 6:\penalty0 215--234, 1968.
\newblock \doi{10.1016/0028-3932(68)90021-3}.

\bibitem[Momennejad et~al.(2017)Momennejad, Russek, Cheong, Botvinick, Daw, and
  Gershman]{momennejad2017}
I.~Momennejad, E.~M. Russek, J.~H. Cheong, M.~M. Botvinick, N.~D. Daw, and
  S.~J. Gershman.
\newblock The successor representation in human reinforcement learning.
\newblock \emph{Nature Human Behaviour}, 1:\penalty0 680--692, 2017.
\newblock \doi{10.1038/s41562-017-0180-8}.

\bibitem[O’Doherty et~al.(2003)O’Doherty, Dayan, Friston, Critchley, and
  Dolan]{odoherty2003}
J.~P. O’Doherty, P.~Dayan, K.~Friston, H.~Critchley, and R.~J. Dolan.
\newblock Temporal difference models and reward-related learning in the human
  brain.
\newblock \emph{Neuron}, 38:\penalty0 329--337, 2003.
\newblock \doi{doi.org/10.1016/S0896-6273(03)00169-7}.

\bibitem[O’Keefe and Dostrovsky(1971)]{okeefe1971}
J.~O’Keefe and J.~Dostrovsky.
\newblock The hippocampus as a spatial map. preliminary evidence from unit
  activity in the freely-moving rat.
\newblock \emph{Brain Res.}, 34:\penalty0 171--175, 1971.
\newblock \doi{10.1016/0006-8993(71)90358-1}.

\bibitem[Pennartz et~al.(2011)Pennartz, Ito, Verschure, Battaglia, and
  Robbins]{pennartz2011}
C.~Pennartz, R.~Ito, P.~Verschure, F.~Battaglia, and T.~Robbins.
\newblock The hippocampal--striatal axis in learning, prediction and
  goal-directed behavior.
\newblock \emph{Trends Neurosci.}, 34:\penalty0 548--559, 2011.
\newblock \doi{10.1016/j.tins.2011.08.001}.

\bibitem[Rebola et~al.(2017)Rebola, Carta, and Mulle]{rebola2017}
N.~Rebola, M.~Carta, and C.~Mulle.
\newblock Operation and plasticity of hippocampal ca3 circuits: implications
  for memory encoding.
\newblock \emph{Nature Publishing Group}, 18:\penalty0 209--221, 2017.
\newblock \doi{10.1038/nrn.2017.10}.

\bibitem[Rushworth et~al.(2011)Rushworth, Noonan, Boorman, Walton, and
  Behrens]{rushworth2011}
M.~F. Rushworth, M.~P. Noonan, E.~D. Boorman, M.~E. Walton, and T.~E. Behrens.
\newblock Frontal cortex and reward-guided learning and decision-making.
\newblock \emph{Neuron}, 70:\penalty0 1054--1069, 2011.
\newblock \doi{10.1016/j.neuron.2011.05.014}.

\bibitem[Schultz(1998)]{schultz1998}
W.~Schultz.
\newblock Predictive reward signal of dopamine neurons.
\newblock \emph{J. Neurophysiol.}, 80:\penalty0 1--27, 1998.
\newblock \doi{doi.org/10.1152/jn.1998.80.1.1}.

\bibitem[Shin et~al.(2019)Shin, Tang, and Jadhav]{shin2019}
J.~D. Shin, W.~Tang, and S.~P. Jadhav.
\newblock Dynamics of awake hippocampal-prefrontal replay for spatial learning
  and memory-guided decision making.
\newblock \emph{Neuron}, 104:\penalty0 1110--1125. e7, 2019.
\newblock \doi{doi.org/10.1016/j.neuron.2019.09.012}.

\bibitem[Silver et~al.(2017)Silver, Schrittwieser, Simonyan, Antonoglou, Huang,
  Guez, Hubert, Baker, Lai, Bolton, Chen, Lillicrap, Hui, Sifre, van~den
  Driessche, Graepel, and Hassabis]{silver2017}
D.~Silver, J.~Schrittwieser, K.~Simonyan, I.~Antonoglou, A.~Huang, A.~Guez,
  T.~Hubert, L.~Baker, M.~Lai, A.~Bolton, Y.~Chen, T.~Lillicrap, F.~Hui,
  L.~Sifre, G.~van~den Driessche, T.~Graepel, and D.~Hassabis.
\newblock Mastering the game of go without human knowledge.
\newblock \emph{Nature}, 550:\penalty0 354--359, 2017.
\newblock \doi{10.1038/nature24270}.

\bibitem[Stachenfeld et~al.(2017)Stachenfeld, Botvinick, and
  Gershman]{stachenfeld2017}
K.~L. Stachenfeld, M.~M. Botvinick, and S.~J. Gershman.
\newblock The hippocampus as a predictive map.
\newblock \emph{Nat. Neurosci.}, 7:\penalty0 1951, 2017.
\newblock \doi{10.1038/nn.4650}.

\bibitem[Suri(2002)]{suri2002}
R.~E. Suri.
\newblock Td models of reward predictive responses in dopamine neurons.
\newblock \emph{Neural networks}, 15:\penalty0 523--533, 2002.
\newblock \doi{doi.org/10.1016/S0893-6080(02)00046-1}.

\bibitem[Sutton and Barto(2018)]{sutton2018}
R.~S. Sutton and A.~G. Barto.
\newblock 1.7. early history of reinforcement learning.
\newblock In \emph{Reinforcement Learning}, pages 13--22. MIT Press, 2018.

\bibitem[Vertes and Sahani(2019)]{vertes2019}
E.~Vertes and M.~Sahani.
\newblock A neurally plausible model learns successor representations in
  partially observable environments, 2019.

\bibitem[Vinyals et~al.(2019)Vinyals, Babuschkin, Czarnecki, Mathieu, Dudzik,
  Chung, Choi, Powell, Ewalds, and Georgiev]{vinyals2019}
O.~Vinyals, I.~Babuschkin, W.~M. Czarnecki, M.~Mathieu, A.~Dudzik, J.~Chung,
  D.~H. Choi, R.~Powell, T.~Ewalds, and P.~Georgiev.
\newblock Grandmaster level in starcraft ii using multi-agent reinforcement
  learning.
\newblock \emph{Nature}, 575:\penalty0 350--354, 2019.
\newblock \doi{10.1038/s41586-019-1724-z}.

\end{thebibliography}

 


\end{document}